\documentclass[a4paper,11pt]{article}
\usepackage{pos}
\usepackage[separate-uncertainty, 
	exponent-product = \cdot,
	per-mode=fraction]{siunitx}
\DeclareSIUnit{\sample}{Sa}
\DeclareSIUnit{\cps}{cps}
\DeclareSIUnit{\year}{yr}
\DeclareSIUnit{\hour}{hr}
\usepackage{circuitikz}[europeanvoltage, betterproportions]
\usepackage{graphicx}
\usepackage{subcaption}
\usepackage{lipsum}
\usepackage{float}
\usepackage{natbib}

\title{Simulations for direct dark matter searches
using\\ ALPS II's TES detection system}
\ShortTitle{Simulations for DDM searches
using ALPS II's TES detection system}

\author*[a]{Christina Schwemmbauer}
\author[b]{Yonit Hochberg}
\author[c]{Katharina-Sophie Isleif}
\author[a]{Friederike Januschek}
\author[d]{Benjamin V. Lehmann}
\author[a]{Axel Lindner}
\author[e]{Manuel Meyer}
\author[f]{Gulden Othman}
\author[a]{José Alejandro Rubiera Gimeno}

\affiliation[a]{Deutsches Elektronen-Synchrotron DESY,
  Notkestr. 85, 22607 Hamburg, Germany}

\affiliation[b]{Racah Institute of Physics, Hebrew University of Jerusalem,
Edmond J. Safra Campus Givat Ram, Jerusalem, Israel}

\affiliation[c]{Helmut-Schmidt-University, Holstenhofweg 85, 22043 Hamburg, Germany}

\affiliation[d]{Center for Theoretical Physics, Massachusetts Institute of Technology, Cambridge, MA 02139, USA}

\affiliation[e]{CP3-Origins, University of Southern Denmark, Campusvej 55, 5230 Odense, Denmark}

\affiliation[f]{Institut für Experimentalphysik, Universität Hamburg, Luruper Chaussee 149, 22761 Hamburg, Germany}

\emailAdd{christina.schwemmbauer@desy.de}
\emailAdd{yonit.hochberg@mail.huji.ac.il}
\emailAdd{katharina-sophie.isleif@desy.de}
\emailAdd{friederike.januschek@desy.de}
\emailAdd{bvl@mit.edu}
\emailAdd{axel.lindner@desy.de}
\emailAdd{manuel.meyer@desy.de}
\emailAdd{gulden.othman@desy.de}
\emailAdd{jose.rubiera.gimeno@desy.de}

\abstract{Transition Edge Sensors (TES) are superconducting microcalorimeters that can be used for single-photon detection with extremely low backgrounds. When they are within their superconducting transition region, small temperature fluctuations---like the energy deposited by single photons---lead to large resistance variations. These variations can be measured using Superconducting Quantum Interference Devices (SQUIDs). This technology is planned to be used as a single-photon detector for later runs of the ALPS II experiment, a light-shining-through-walls experiment at DESY Hamburg, searching for Axion-Like Particles (ALPs), which are possible Dark Matter (DM) candidates. Due to the very low dark count rates in our setup, our TES system might be viable for direct DM searches at sub-MeV masses through electron-scattering of DM in the superconducting chip, as well. Simulations concerning background rejection and calibration methods demonstrate the needed sub-eV sensitivity already.}
\FullConference{1st General Meeting and 1st Training School of the COST Action COSMIC WSIPers (COSMICWISPers)\\
 5-14 September, 2023\\
Bari and Lecce, Italy\\}

\begin{document}
\maketitle

\section{Dark Matter searches with TES}

    Many direct Dark Matter (DM) experiments focus on the search for WIMPs (Weakly Interacting Massive Particles) exploiting DM scattering on heavy nuclei and measuring their recoil energy \cite{feng:2010}. However, WIMPs have not been found so far and experiments will reach the neutrino fog soon \cite{bertone:2018, Akerib2022}, which might exhaust these approaches in the near future. To explore different parameter spaces at lower masses DM-electron scattering can be explored, which poses an alternative to the scattering off of nucleons. When using the rule of thumb from Ref.~\cite{Hochberg2019}, one finds a relation between the mass of the projectile DM particle and the transferred energy in electron-scattering experiments. Correspondingly, a possible avenue for DM-electron scattering are superconducting targets with sub-eV sensitivity. These would be able to probe sub-MeV DM masses, including well-motivated DM models as well \cite{Knapen2017}.

    In this work, we explore the possibility of using a Transition Edge Sensor (TES) for sub-MeV DM detection using DM-electron scattering. This could be proof-of-principle for new types of DM detectors. For this purpose, we propose to use the TES setup of the ALPS II (Any Light Particle Search) experiment, equipped with TES sensors provided by NIST (USA) and SQUIDs provided by PTB (Germany). 

    The concept of directly searching for DM with superconductors has already been shown with SNSPDs (Superconducting Nanowire Single Photon Detectors) \cite{Hochberg2019, Hochberg2022}. Similarly, TES could be used as a simultaneous target and readout for impinging DM particles. SNSPDs function like Geiger counters, where nearly no information from the initial energy deposition is preserved. The only particle parameter that can be derived from an SNSPD signal, is the energy surpassing the device's threshold for breaking Cooper pairs. When using TES for the same purpose, more information is preserved since the whole pulse shape of a triggered signal is stored and subsequently analyzed. Furthermore, TES might even be able to measure DM signals at lower energy thresholds in comparison to SNSPDs \cite{Hochberg2015}.

\begin{figure}[ht]
        \centering
        \includegraphics[width = 0.66\textwidth,trim=3cm 2cm 0.5cm 0.5cm, clip]{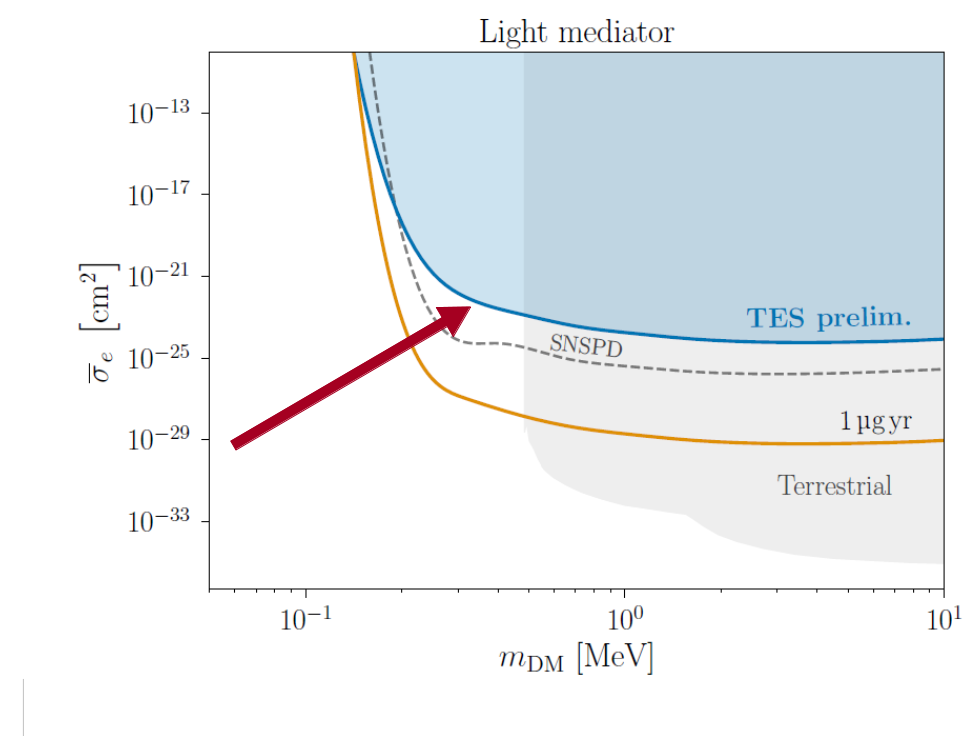}
        \caption{Terrestrial parameter space for DM-electron scattering with a light mediator in the $\sim$MeV mass range dependent on the DM-electron scattering cross section $\Bar{\sigma}_e$. The blue area shows the preliminary projection (\SI{20}{\nano\gram\hour}) for our TES based on intrinsics measurements (for ALPS II - not optimized). The dotted gray line describes the parameter space covered by SNSPD measurements (\SI{774}{\nano\gram\hour}) with the projection for a larger exposure of SNSPDs in orange (from \cite{Hochberg2022}). The shaded gray area represents other terrestrial constraints.}
        \label{fig::TES_projection}
    \end{figure}

    The tungsten TES setup that will be used in later runs of the ALPS II experiment, aims to detect single photons from (Axion-Like-Particle) ALP-photon oscillations of a rate down to $\SI{e-5}{\second^{-1}}$ \cite{Bähre2013, Shah2022}. For this setup, routine intrinsic measurements are conducted to explore the background for the ALPS II measurements. These intrinsic background measurements are carried out without any optical fibers connected, such that the TES is isolated inside of a cryostat. Therefore, these measurements are suitable for direct DM searches as well. Figure \ref{fig::TES_projection} shows a comparison of the expected performance of our TES setup (based on previous intrinsic measurements) with the SNSPD results from \cite{Hochberg2022}.

\begin{figure}[t]
        \centering
        \begin{circuitikz}
            \draw (0,2) to[battery1,v_=$V_\mathrm{bias}$] (0,0);
            \draw (0,2) to[R, l=$R_L$] (2,2);
            \draw (2,2) to[L, l_=$L$] (2,0);
            \draw (0,0) to[vR, l_={$R_\mathrm{TES}(T,I)$}] (2,0);
            \draw (0,0) -- (0,-0.5) node[ground]{};
            \draw (3.5,2) to[squid, l=SQUID, label position = {(2.5,0.5)}] (3.5,0);
            \draw (3.5,2) to[short,-o](3.5,2) node[above] {$V_\mathrm{out}$};
            \draw (3.5,0) -- (3.5,0) node[ground]{};
        \end{circuitikz}
        \caption{Simplified circuit of a TES with SQUID readout. The TES is biased through a voltage source $V_\mathrm{bias}$ and its resistance $R_\mathrm{TES}$ is temperature $T$ and current $I$ dependent. Any magnetic flux variations arising from changing current flowing through the coil with inductance $L$ will be picked up by the SQUID, functioning as a sensitive magnetometer.}
        \label{fig::circuit}
    \end{figure}
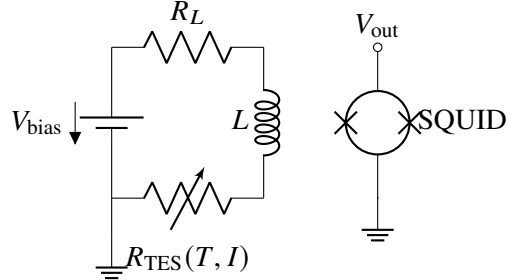

    Generally, TES measure single-photon pulses by keeping the biased superconducting chip at a transition temperature between the superconducting and normal conducting state (\SI{140}{\milli \kelvin} in this case). When a single photon hits the TES, the absorbed energy leads to a slight increase in temperature, which results in a relatively large resistance increase. Through the coupling of the TES chip to SQUIDs (Superconducting Quantum Interference Devices) this change is read out and finally translated to a voltage pulse. The resulting pulse shape of the signal, such as the rise and decay time of the pulse, is primarily dependent on the TES circuit (see Figure \ref{fig::circuit}) \cite{Irwin2005}.
    Other parameters like the pulse height or pulse integral vary depending on the energy deposited in the interaction \cite{Irwin2005}. These parameters are determined by fitting a pulse shape function for \SI{1.165}{\eV} signals (see~Ref.\cite{RubieraGimeno2023}). Signals from DM particles should be similar since both processes rely on the breaking of Cooper pairs \cite{Hochberg2016}. 

\section{Energy Calibration}

    The dependence of the TES response on the energy previously showed a linear behavior of the pulse height for photon energies above \SI{1}{\eV} \cite{NoemiThesis}. Since our TES setup is optimized for \SI{1.165}{\eV} (\SI{1064}{\nano\meter}), we also need to understand the TES response for energies below \SI{1}{\eV} where we expect to reach lower thresholds than the above-mentioned SNSPD device. For this purpose, we will use a setup of butterfly laser diodes ranging from \SI{1.4}{\eV} to \SI{0.6}{\eV} (or \SI{880}{\nano\meter} - \SI{2000}{\nano\meter}). By investigating the TES pulse shapes of single photons of different energies, we want to evaluate the linearity of the energy response of the TES  in this range to improve our analysis for direct DM searches. As a first step, we performed simulations (see Figure \ref{fig::TES_triggers}) of the pulse shape we expect for photons from these laser diodes, which we derive from our well-tested simulation of \SI{1064}{\nano\meter} pulses \cite{RubieraGimeno2023}. We keep the rise and decay times constant while adjusting the pulse height proportional to the expected energy (determined by measurements with optical spectrometers). We plan to compare these simulated pulses and the linear energy dependence with calibration measurements of our TES setup soon.

\section{Electronic noise simulations}\label{sec::simulations}

\begin{figure}
        \begin{minipage}[b]{0.49\linewidth}
            \centering
            \includegraphics[width=\linewidth]{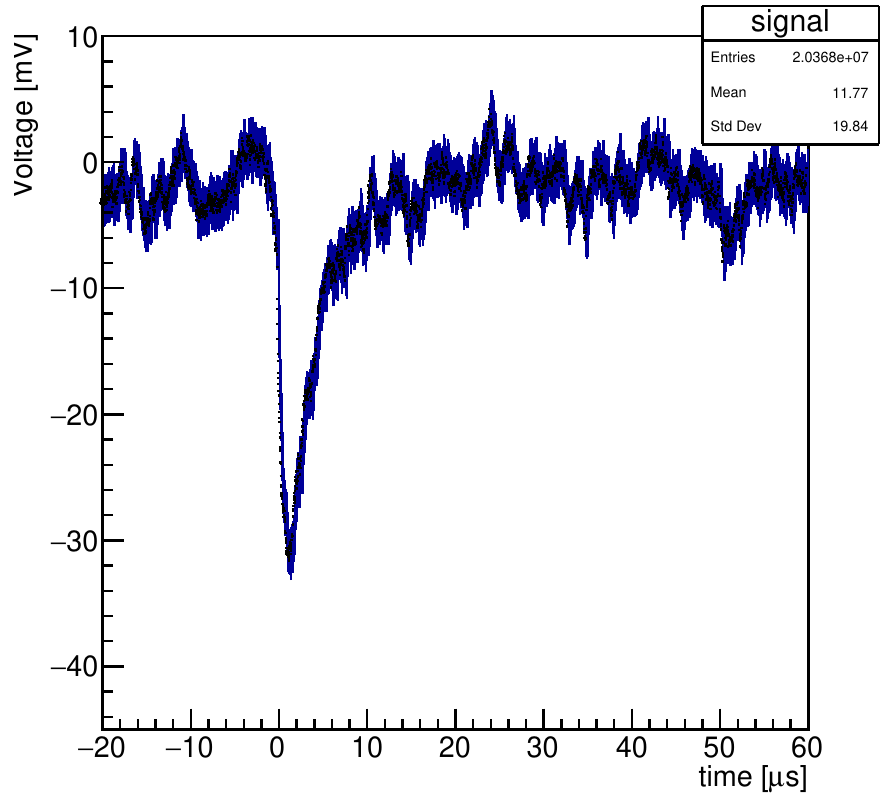}
        \end{minipage}
        \hfill
        \begin{minipage}[b]{0.49\linewidth}
            \centering
            \includegraphics[width=\linewidth]{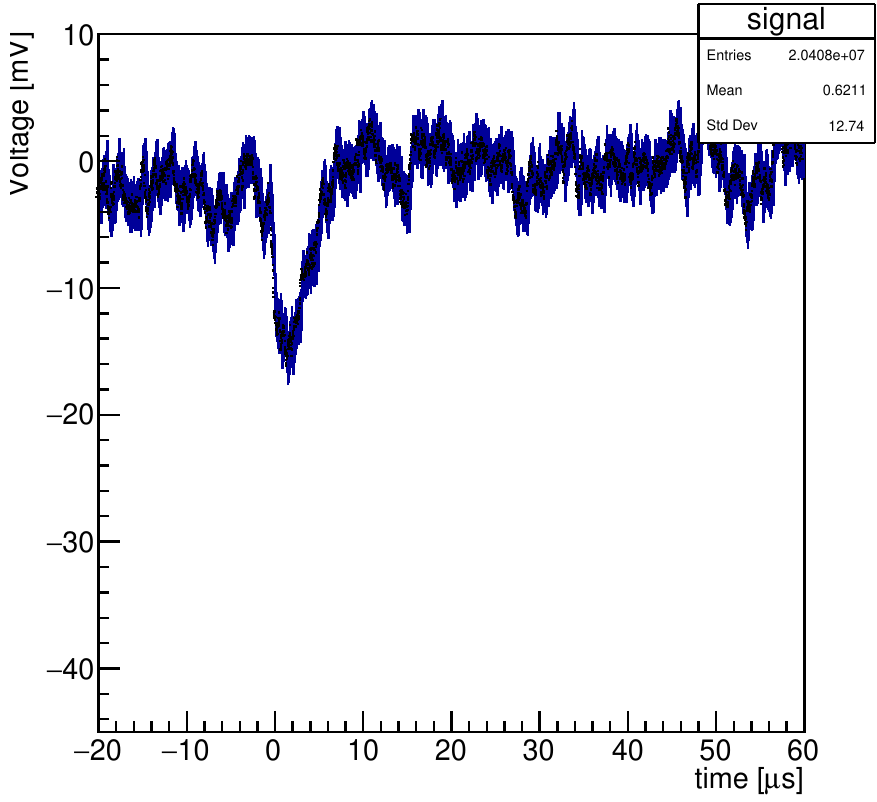}
        \end{minipage}
        \caption{Simulated example TES pulses for \SI{1.165}{\eV} (left) and \SI{0.583}{\eV} (right).}
        \label{fig::TES_triggers}
    \end{figure}

    To explore the viability of the TES for these direct DM searches, we also need to understand how well we can discern electronic noise from lower energy signals. When conducting a measurement, a pre-selection happens via a threshold trigger that needs to be lowered (in comparison to \SI{1.165}{\eV}) to access lower energies. With our existing simulation and analysis framework (for an example of light pulses see Figure \ref{fig::TES_triggers}), we simulate the electronic noise of our detector and mimic the noise component of an intrinsic background measurement with lower triggers.
    When reducing the trigger level for a \SI{500}{\second} simulation by about \SI{30}{\percent}, the rate of triggered background signals increases from no triggers to \SI{0.32\pm0.03}{\hertz}. Therefore, we need to improve background rejection in the subsequent analysis to remove triggered noise background pulses resulting from the reduced signal-to-noise ratio.
    
    For this purpose, we simulate lower energy pulses (\SI{0.583}{\eV}, see Figure \ref{fig::TES_triggers}, right) and use this lower trigger level again. In the subsequent analysis, we identify typical values for fit parameters like pulse height, rise time, and decay time linked to lower energy signals (0.583 eV). These values help us set specific criteria for making precise cuts, which are crucial for isolating background signals. When applying these stringent cuts to the simulated pulses, we reach an acceptance of \SI{56}{\percent} for this energy bin. Subsequently, these cuts are applied to an electronic-noise-only simulation to determine the rate of surviving background pulses. The cut-based analysis reduces the trigger rate from \SI{0.422\pm0.010}{\hertz} (without analysis) to no surviving signals with an upper uncertainty limit of \SI{0.0007}{\hertz} at \SI{95}{\percent} C.L.. This is a promising result that shows the competitiveness of our setup with the above-mentioned SNSPD results with a threshold of \SI{0.73}{\eV}. We aim to verify these results experimentally with dedicated measurement and analysis efforts soon.

    In conclusion, the TES setup for the ALPS II experiment shows promising prospects to be used for direct DM searches as well. With the explored analysis methods for calibration and noise background reduction, our TES setup could pave the way for a new type of direct DM detector for sub-MeV DM searches exploiting DM electron scattering. After confirming our simulation results experimentally, we also plan to improve our setup and possibly optimize it for direct DM searches e.g. by modifying the TES detector modules.
  
\section*{Acknowledgments}
    We would like to thank our ALPS collaborators, the National
    Institute of Standards and Technology (NIST), USA, for the TES devices and Physikalisch-Technische Bundesanstalt (PTB) Berlin and the Humboldt University Berlin, Germany, for the SQUID modules and helpful advice. M.M. acknowledges the support from the European Research Council (ERC) under the European Union's Horizon 2020 research and innovation program Grant agreement No. 948689 (AxionDM) and the support from the Deutsche Forschungsgemein-schaft (DFG, German Research Foundation) under Germany’s ExcellenceStrategy – EXC 2121 “Quantum Universe” – 390833306. This article is based upon work from COST Action COSMIC WISPers CA21106, supported by COST (European Cooperation in Science and Technology).
\bibliographystyle{myJHEP} 
\bibliography{skeleton}
\end{document}